\documentclass{Interspeech2024}

\interspeechcameraready 

\usepackage{times}
\usepackage{latexsym}

\usepackage[T1]{fontenc}

\usepackage[utf8]{inputenc}

\usepackage{microtype}

\usepackage{inconsolata}
\usepackage{url}
\usepackage{booktabs}
\usepackage{amsfonts}
\usepackage{nicefrac}
\usepackage{microtype}      
\usepackage{subcaption}
\usepackage[ruled]{algorithm2e}
\SetAlFnt{\small}
\usepackage{amsmath}
\usepackage{bm}
\usepackage{bbold}
\usepackage{graphicx}
\usepackage{multirow}
\usepackage{textcomp}
\usepackage{wrapfig}
\usepackage{subcaption}
\usepackage{xspace}
\usepackage{interval}
\usepackage{bm,upgreek}
\usepackage{colortbl}
\usepackage{enumitem}
\usepackage{pifont}
\usepackage{multirow,makecell}
\usepackage{xcolor}

\title{Domain Adaptation for Contrastive Audio-Language Models}

\name[affiliation={1,2}]{Soham}{Deshmukh}
\name[affiliation={1}]{Rita}{Singh}
\name[affiliation={1}]{Bhiksha}{Raj}

\address{
  $^1$Carnegie Mellon University, 
  $^2$Microsoft}
\email{\{sdeshmuk,rsingh,bhikshar\}@andrew.cmu.edu}


\begin{document}

\setlength{\belowdisplayskip}{4pt} \setlength{\belowdisplayshortskip}{4pt}
\setlength{\abovedisplayskip}{4pt} \setlength{\abovedisplayshortskip}{4pt}

\maketitle
\begin{abstract}
Audio-Language Models (ALM) aim to be general-purpose audio models by providing zero-shot capabilities at test time. The zero-shot performance of ALM improves by using suitable text prompts for each domain. The text prompts are usually hand-crafted through an ad-hoc process and lead to a drop in ALM generalization and out-of-distribution performance. Existing approaches to improve domain performance, like few-shot learning or fine-tuning, require access to annotated data and iterations of training. Therefore, we propose a test-time domain adaptation method for ALMs that does not require access to annotations. Our method learns a domain vector by enforcing consistency across augmented views of the testing audio. We extensively evaluate our approach on 12 downstream tasks across domains. With just one example, our domain adaptation method leads to 3.2\% (max 8.4\%) average zero-shot performance improvement. After adaptation, the model still retains the generalization property of ALMs.  
\end{abstract}
\section{Introduction}
Audio-Language Models (ALMs) use language as a mode of supervision to learn general-purpose audio representations \cite{Elizalde2023NaturalLS, wu2022large, pengi, gong2023listen}. Once pretrained, the ALMs can be used for multiple downstream tasks. For instance, the widely used paradigm of CLAP \cite{wu2022large, mei2023wavcaps} contrastively trains an audio and text encoder on millions of audio-text pairs and at test time shows impressive generic zero-shot performance. Then the model is employed in a zero-shot setup across various domains from Sound Event Classification to Speech Emotion Recognition. For each domain, human input in crafting text prompts helps in improving the zero-shot performance of ALMs. 

The human prompt engineering improves the ALM's performance on the specific domain. For example, manual prompt engineering \cite{elizalde2022clap} from \textit{``\{class\}"} to \textit{``this is a sound of \{class\}"} leads to a relative 2\% improvement in performance on audio event datasets. The text prompts are designed for each domain using an ad-hoc experimental process. This process generally requires knowledge of the ALM's training data which may or may not be available to the public. Existing approaches improve domain performance by Few-Shot Learning (FSL). This involves learning prompts \cite{li2023audio} or adapters \cite{liang2023adapting} using supervised training on annotated target data. The annotation process is expensive and may not be feasible for deployed models. FSL methods also lead to a loss in generalization of base ALM and poor out-of-distribution performance \cite{10097117,liang2023adapting}. 

Parallely, in the vision domain, there have been two model adaptation frameworks: FSL and Test-Time Training \cite{sun2020test}. The examples for FSL are CoOp \cite{zhou2022learning} and CoCoOp \cite{zhou2022learning}, both of which require access to labels. On the other hand, Test-Time Training \cite{sun2020test} focuses on using test examples and auxiliary tasks for full or partial model \cite{NEURIPS2022_bcdec1c2, wang2021tent} and prompt updates \cite{shu2022test}. Test-time tuning is a promising direction because it does not make assumptions about target data distribution and does not require labeled data. However, the existing Test-Time Training methods require either an additional self-supervision branch, multiple test-time examples, or suffer from high computation costs due to parameter updates. Specifically for audio, there has not been an exploration of test-time adaptation for ALMs, let alone compute and parameter-efficient test-time updates.  

In this work, we propose a domain adaptation method for Contrastive ALMs that is performed at test-time using unlabeled audio. Our method involves learning a domain vector, which helps the model adjust to new audio domains by enforcing consistent prediction across augmented views of audio. This process is divided into three stages: Augment: The test audio is subject to various augmentations to simulate different listening conditions. Alongside this, a domain vector is incorporated into the model's text processing component; b) Combine: The model makes predictions based on the augmented audio and the modified text embeddings. These predictions are then averaged to produce a more stable and generalized output, and c) Optimize: Finally, the averaged prediction probabilities are used to compute self-entropy to fine-tune the domain vector, ensuring that the model's predictions are adapted for the domain. With only one unlabelled audio, we see an average 3.2\% (max 8.4\%) zero-shot performance improvement across 12 downstream tasks, while retaining the generalization property of ALMs.

\begin{figure*}[!t]
\vspace{-15pt}
\centering
\resizebox{0.9\linewidth}{!}{
\includegraphics{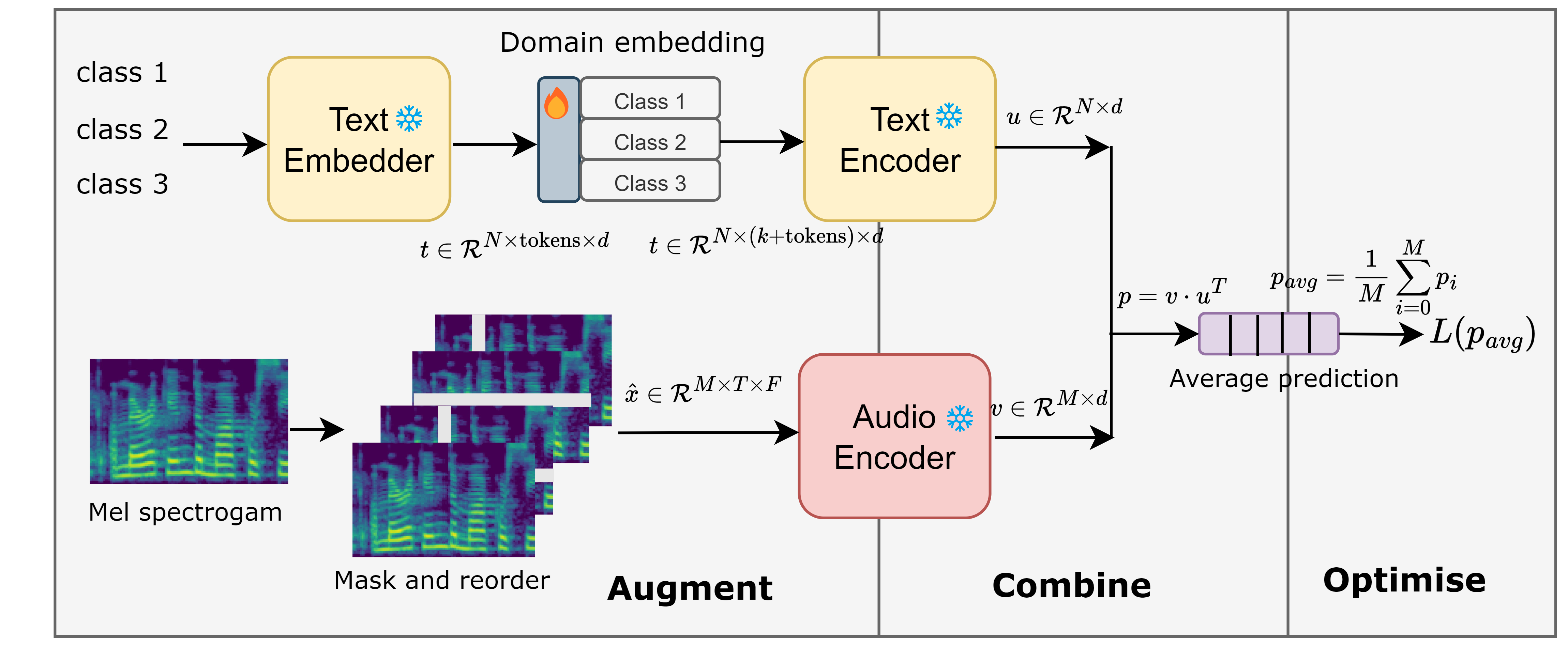}
}
\caption{Our method takes a single input audio at test time and optimizes a domain embedding using an entropy-based loss function. The method does not require labels and enforces consistent prediction across augmented views.}
\vspace{-15pt}
\label{fig:intro}
\end{figure*}

\section{Background}
\subsection{Contrastive Audio-Language Models} \label{sec: clap}
This section provides a generic overview of Contrastive Audio-Language models in literature. We will commonly refer to them as ``model". The model has a two-tower structure consisting of audio and text branches. The two branches are pretrained using contrastive learning \cite{radford2021learning}. After pretraining, the model can be used for zero-shot classification. To illustrate, the user provides an audio file and the $N$ classes to classify the audio into. The audio file is converted into a mel-spectrogram ($x \in \mathbb{R}^{T \times F}$) and passed to an audio encoder to produce audio embeddings $v \in \mathbb{R}^{1 \times d}$. On the text side, $N$ classes are tokenized and embeddings are obtained for each token. Then the token embeddings are passed through a text encoder to produce text embeddings $u \in \mathbb{R}^{d \times N}$. Both embeddings are projected into a common embedding space using two independent projection layers. After projection, the dot product is computed between the two embeddings. Generally, an activation function like Softmax or sigmoid is applied to obtain a valid probability distribution across $N$ classes of interest. 

\subsection{Human-crafted text-prompts}\label{sec: human crafted prompts}
The contrastive models like CLAP \cite{elizalde2022clap} append prompts to classes for better zero-shot performance. The prompts are hand-crafted for the domain the model is being tested on. For example, a prompt of "this emotion is " is used for Speech Emotion Recognition while the prompt of "this is the sound of " is used for Sound Event Classification. The choice of text prompt for each task affects performance \cite{elizalde2022clap,deshmukh2024pam}. From the prompt experiment, we hypothesize that: \textit{though audio-language models have an adequate understanding of audio concepts, we don't have methods to elicit this information from the models}. The manual method of hand-crafting prompts is cumbersome and requires domain knowledge. Moreover, the handcrafted prompts don't take into account intra-task variance resulting from domain changes. For example, zero-shot inference for the task of sound event classification can provide different performances when evaluated in an urban city, a domestic household versus in a rainforest. This leads us to believe an automatic unsupervised adaptation method for zero-shot inference of audio-language can help improve performance. 

\section{Methodology}\label{sec: method}
Our approach can be summarized as learning a domain vector $a$ that minimizes the entropy of prediction probability obtained from multiple audio augmentations for test audio:
$$\min_a - p\log p $$
where $p$ is the average augmentation probability distribution: $\frac{1}{M}\sum_{i=1}^M p(y_i|z_i))$, $z_i$ is the $i^{th}$ augmentation for test audio $x$, $y_i$. The method is depicted in Figure \ref{fig:intro} and is divided into three parts: Augment, Combine, and Optimize.

\subsection{Augment}\label{sec: augment}
The audio file is converted to a mel spectrogram ($x \in \mathbb{R}^{T \times F}$). Rather than directly passing $x$ to the audio encoder, it is first augmented using multiple random yet controlled augmentations. The augmentation is masking-based and inspired by SpecAug \cite{park2019specaugment}. The four main augmentations performed are:\\ 
\textbf{Time Masking (TM):} Random portions of spectrograms are masked along the time axis. The width of the mask ranges from 2 to 128. The number of masked stripes ranges from 2 to 24. \\
\textbf{Frequency Masking (FM):} Random portions of spectrograms are masked along the freq. axis. The width of the mask ranges from 2 to 32. The number of masked stripes ranges from 2 to 24. \\
\textbf{Time \& Frequency Masking (TFM):} Random portions of spectrograms are masked along both the frequency and time axis. This is the combination of time masking followed by frequency masking. \\
\textbf{Time Reorder (TR):} Audio-language models compress all audio information into a single embedding which leads to loss of temporal information \cite{wu2023audio}. Therefore, we explicitly add a time-reorder augmentation (details in appendix). \\
\noindent All the augmentation along with the original input $x$ is concatenated to form a batch of inputs: 
$$\hat{x} = [\text{TM}(x), \text{FM}(x), \text{TFM}(x), \text{TR}(x), x]$$
where $\hat{x} \in \mathbb{R}^{M \times T \times F}$ where $M$ is number of augmentations plus one original spectrogram. The augmented and batch of spectrograms $\hat{x}$ is passed through the audio encoder and linear projection to obtain audio embedding $v \in \mathbb{R}^{M \times d}$. 

For the text side, the $N$ classes are first tokenized and converted to token-level embeddings. Then a learnable domain vector is concatenated with the tokenized embeddings. The learnable domain vector is randomly initialized. If the token-level text embedding is $t \in \mathbb{R}^{N \times \text{tokens} \times d}$ and is concatenated with the learnable vector of $k$ tokens, the output embedding is $t \in \mathbb{R}^{N \times (\text{tokens}+k) \times d}$. Therefore, the method optimizes only one vector for the domain and the one vector does not vary per class. The $t$ embedding is passed through a text encoder to produce text embeddings $u \in \mathbb{R}^{d \times N}$. For our discussion, we consider the independent projection layers as a part of audio and text encoder respectively. 

\begin{table*}[ht]
\footnotesize
\center 
\begin{tabular}{c|c|ccc|c|c|c} \hline
& \multicolumn{1}{c|}{\makecell{ Average $\uparrow$}} & \multicolumn{3}{c|}{Sound Event Classification $\uparrow$} & \multicolumn{1}{c|}{\makecell{Vocal Sound \\ Classification $\uparrow$}} & \multicolumn{1}{c|}{\makecell{Surveillance \\ Sound Classif.$\uparrow$}} & \multicolumn{1}{c}{\makecell{Acoustic Scene \\ Classification$\uparrow$}}\\ \hline
\makecell{Model} & \makecell{Average} & ESC50 & US8K & \makecell{DCASE17 \\ Task 4} & \makecell{Vocal\\Sound} & \makecell{SESA} &\makecell{TUT 2017} \\ \hline 
Zero-Shot & 62.93 & 93.90 & 82.30 & 46.60 & 79.97 & 64.95 & 53.80 \\ 
One audio & 64.94 & 93.35 & \textbf{85.26} & 50.96 & 82.14 & 73.30 & 54.19 \\ 
Five audio & \textbf{65.92} & \textbf{95.05} & 85.21 & \textbf{52.30} & \textbf{82.40} & \textbf{74.35} & \textbf{54.38} \\ \hline
\end{tabular}
\smallskip
\center
\begin{tabular}{c|cc|cc|cc} \hline
& \multicolumn{2}{c|}{Music Classification $\uparrow$} & \multicolumn{2}{c|}{Instrument Classification $\uparrow$} & \multicolumn{2}{c}{\makecell{ Speech Emotion \\ Classification$\uparrow$}} \\ \hline
\makecell{Model} & \makecell{GTZAN\\Music Speech} & \makecell{GTZAN\\Genres} & \makecell{Beijing\\Opera} & \makecell{NS Instr.\\family} & \makecell{CRE\\MA-D} & \makecell{RAV\\DESS} \\ \hline
Zero-Shot & 99.20 & 58.40 & 46.60 & 68.00 & 30.00 & 31.54 \\ 
One audio & 99.21 & 61.00 & 47.45 & 68.28 & 29.92 & \textbf{34.25} \\ 
Five audio & \textbf{1.00} & \textbf{63.20} & \textbf{50.42} & \textbf{69.23} & \textbf{31.27} & 33.19 \\ \hline
\end{tabular}
\caption{\label{table: encoder results}
 Zero-shot and Test-time adaptation performance on 12 downstream tasks. The adaptation method uses one randomly chosen unlabelled audio example and five randomly chosen unlabelled audio examples at test time. The Table reports the average numbers across 5 runs. The metric is accuracy and higher is better for all tasks. Average $\uparrow$ is the average performance across 12 tasks} \vspace{-0.2in}
\end{table*}

\subsection{Combine}
The audio embedding $v \in \mathbb{R}^{M \times d}$ and text embeddings $u \in \mathbb{R}^{d \times N}$ are used to compute a dot product:
$$p = v \cdot u^{T}$$
where $p \in \mathbb{R}^{M \times N}$. We use softmax to convert $p$ into a valid probability distribution across classes $N$. After softmax, $p$ is the probability of each class for a total of $M$ augmented views of the audio. The probability distribution $p$ is averaged along the different augmentations $M$ to get:
$p_{avg} = \frac{1}{M}\sum_{i=1}^M p_i$
where $p_{avg}$ in $\mathbb{R}^{N}$

\subsection{Optimise}\label{sec: optimise}
The $p_{avg}$ is the average of augmented probability distributions over $N$ classes. To learn the domain vector, we have to use a loss function. As the labels are not known, the constructed loss function has to be a self-supervision or unsupervised loss. Therefore, we choose self-entropy as the loss function. By minimizing self-entropy, we enforce consistency across the predictions of augmented views. Therefore, the loss function is:
$$\mathcal{L} = - p_{\text{avg}} \log p_{\text{avg}}$$
The loss $\mathcal{L}$ is calculated using a single test example and optimizes a parameter of size: $k \times d$, where $k$ is the number of tokens.  
\section{Experiments} 
\label{sec:exp}
\subsection{Downstream tasks}\label{subsec:downstream tasks}\vspace{-0.05in}
We benchmark our method on 12 downstream tasks \cite{turian2022hear} across the domains of Sound Event Classification, Acoustic Scene Classification, Vocal Sounds, Music, Surveillance, and Speech Emotion Recognition. The datasets used are: ESC50 \cite{esc50}, UrbanSound8K \cite{UrbanSound}, DCASE2017 Task4\cite{mesaros2017dcase}, TUT 2017, GTZAN Music Speech \cite{tzanetakis_essl_cook_2001}, GTZAN Genres \cite{tzanetakis_essl_cook_2001}, Beijing Opera Percussions \cite{turian2022hear}, CREMA-D \cite{cao2014crema}, RAVDESS \cite{ravdess}, Vocal Sound \cite{vocalsound}. The datasets have varied audio duration, classes, files, and setups. For example, the audio duration ranges from 3 to $\geq$ 35 seconds, classes range from binary to 50 classes, audio files per data range from 120 to 305k files.

\subsection{Implementation details}
\textbf{Contrastive ALM.} We use the SoTA Contrastive Audio-Language Models \cite{Elizalde2023NaturalLS} for our experiments. The audio encoder is HTSAT\cite{chen2022hts} and the text encoder is a modified GPT2 \cite{radford2019language, ac3training}. The audio is sampled at 44.1 kHz and converted to log Mel spectrograms with 64 Mel bins, a hop size of 320, and a window size of 1024 in the range of 50-14000 Hz range. We use the zero-shot setup of ALM as the baseline. \\
\textbf{Test-Time Adaptation.}  We use a single token of dimension 768 as the learnable domain vector. Relating to Section \ref{sec: optimise}, this implies $k$ is 1 and $d$ is 768. The optimizer used is Weighted Adam \cite{adam} with a learning rate 5e-2.
\section{Results}
\noindent The Results are organized as follows: Section \ref{results: main results} covers the domain adaptation results, Section \ref{results: augmentations} describes the effect of number of augmentations, Section \ref{sec: cross domain gen} studies cross-domain generalization, and Section \ref{results: limitations} discusses limitation of the proposed method.

\subsection{Test-time domain adaptation} \label{results: main results}
The main results are shown in Table \ref{table: encoder results}. As this is the first test-time adaptation method for audio, the zero-shot performance of the model to be adapted (CLAP) is considered baseline performance. The adaptation technique using only one example, provides an average relative improvement of 3.18\% in zero-shot performance. When 5 samples are used for adaptation, the improvement increases to ~4.7\%. 

\begin{table*}[]
\footnotesize
\center
\addtolength{\tabcolsep}{-0.1em}
\begin{tabular}{l|l|llllllllllll|r} \hline
Dataset & ZS & ESC50 & US8K & D17T4 & SESA & GT MS & Genre & Opera & CREM & RAVD & Vocal & TU17 & NSyn & \multicolumn{1}{l}{Avg.} \\ \hline
ESC50 & 93.90 & \cellcolor{blue!25}93.35 & 81.84 & 46.74 & 72.16 & 100.0 & 60.60 & 40.56 & 26.19 & 31.16 & 81.18 & 54.26 & 67.46 & 62.96 \\
US8K & 82.30 & 90.80 & \cellcolor{blue!25} 85.26 & 49.93 & 67.35 & 97.66 & 54.40 & 44.07 & 23.57 & 32.47 & 76.74 & 45.05 & 62.71 & 60.83 \\
D17T4 & 46.60 & 90.60 & 84.72 & \cellcolor{blue!25} 50.96 & 69.40 & 96.88 & 54.70 & 44.92 & 26.86 & 26.96 & 76.02 & 45.48 & 62.81 & 60.86 \\
SESA & 64.95 & 92.40 & 81.45 & 45.26 & \cellcolor{blue!25} 73.30 & 99.22 & 56.30 & 33.47 & 28.02 & 32.22 & 81.87 & 51.60 & 65.26 & 61.70 \\
GT MS & 99.20 & 90.75 & 85.01 & 49.70 & 65.64 & \cellcolor{blue!25} 99.21 & 54.70 & 36.02 & 24.96 & 27.59 & 75.91 & 45.31 & 60.42 & 59.60 \\
Genre & 58.40 & 92.55 & 82.38 & 46.15 & 72.60 & 98.44 & \cellcolor{blue!25} 61.00 & 36.44 & 29.54 & 32.14 & 82.21 & 49.75 & 64.05 & 62.27 \\
Opera & 46.60 & 92.40 & 82.56 & 46.74 & 73.01 & 99.22 & 56.59 & \cellcolor{blue!25} 47.45 & 27.61 & 32.30 & 82.46 & 53.58 & 64.50 & 63.20 \\
CREM & 30.00 & 92.60 & 81.92 & 46.07 & 73.58 & 99.22 & 56.69 & 36.02 & \cellcolor{blue!25} 29.92 & 32.38 & 82.18 & 52.59 & 64.99 & 62.35 \\
RAVD & 31.54 & 92.50 & 81.37 & 45.41 & 74.46 & 99.22 & 59.10 & 37.71 & 29.39 & \cellcolor{blue!25} 34.25 & 82.04 & 52.16 & 65.23 & 62.74 \\
Vocal & 79.97 & 92.75 & 81.88 & 46.07 & 74.09 & 99.22 & 58.19 & 36.44 & 28.29 & 32.22 & \cellcolor{blue!25} 82.23 & 51.91 & 64.97 & 62.36 \\
TU17 & 53.80 & 92.40 & 81.39 & 45.33 & 73.92 & 99.22 & 56.39 & 33.47 & 28.02 & 32.26 & 81.87 & \cellcolor{blue!25} 54.19 & 65.21 & 61.97 \\
NSyn & 68.00 & 92.50 & 81.16 & 45.56 & 73.58 & 99.22 & 55.75 & 34.32 & 27.32 & 32.22 & 81.51 & 51.36 & \cellcolor{blue!25} 68.28 & 61.90 \\ \hline
Avg. & 62.93 & 92.13 & 82.58 & 46.99 & 71.92 & 90.64 & 57.03 & 38.41 & 27.47 & 31.51 & 80.52 & 50.61 & 64.66 & 61.21\\ \hline
\end{tabular}
\caption{\label{appendix: cross-domain generalization} Cross-domain generalization. The rows indicate the dataset from which one sample is used to learn the domain vector. The columns indicate the target dataset used for evaluation. The ZS column is the baseline zero-shot performance of CLAP. The blue diagonal indicates the performance when the sampled audio used to learn the domain vector, comes from the target domain.} \vspace{-0.2in}
\end{table*}

For some downstream tasks like SESA, the improvement is 8.35\% and 9.4\% for 1 and 5 example adaptation respectively. The SESA task consists of 16 kHz audio with up to 33 seconds with classes Casual, Gunshot, Explosion, and Siren (also contains alarms). However, the audio is hard to distinguish even for humans, for example, the audio marked Casual could be a fireworks, factory machine, or thunder which might perceptual sound like other threat classes. The adaptation method forces consistency among predictions for different augmented views, i.e. it learns a vector that makes the model robust to different views of audio in the wild. We hypothesize that this is helpful for scenarios where the sound is perceptually similar and requires more fine-grained prompts to elicit the right information from audio-language models for classification. The downstream task of DCASE2017Task4 has similar perceptually hard-to-distinguish sounds, where the adaptation method helps to improve zero-shot performance by 4.36\% and 5.7\%. 

For the downstream task of Speech Emotion Recognition (SER), specifically dataset CREMAD, the adaptation method leads to a drop in zero-shot performance (-0.08\%) when provided with one example. CLAP is not trained on speech data and performs poorly on SER in a zero-shot setup. For example, the zero-shot performance is 30\%, the random performance is 17\% and the supervised performance is 75.2\%. The CLAP zero-shot performs poorly when the audio encoder has not learned relevant features for the tasks. In such cases, our method leads to minor or no improvements, such as in Speech Emotion Recognition and Keyword Word Spotting.

\subsection{Effect of Augmentations} \label{results: augmentations}
For the results in Table \ref{table: encoder results}, the audio files are augmented 50 times using augmentation strategies described in \ref{sec: augment}. The larger the number of augmentation, the more views the adaptation method can utilize to improve performance. To study the effect of augmentation, we perform an ablation study where only an audio file undergoes half the augmentation leading to a smaller number of augmented views. The comparison results are shown in Table \ref{table:aug}. On average, the larger number of augmentation helps when either one example or five examples are used for adaptation. Therefore, even with 25 augmented views, the domain adaptation improves over a zero-shot baseline. The higher number of augmentations (views) helps in improving performance on harder tasks. Therefore, with fewer augmentation, we see a drop in performance on SESA where audio is perceptually hard to classify. The number of augmentations used is a hyperparameter and can tuned and determined based on the compute power available at inference.

\begin{table}[!ht]
\center
\footnotesize
\begin{tabular}{lcc}
\hline
 & One test example & Five test example \\ 
\hline
25 Augmentations & 63.51 & 63.77 \\
50 Augmentations & \textbf{64.94} & \textbf{65.92} \\ \hline
\end{tabular}
\caption{Effect of number of augmentations on domain adaptation performance for one example and five example setup. \label{table:aug} \vspace{-0.2in}}
\end{table}

\subsection{Cross-domain generalization} \label{sec: cross domain gen}
\vspace{-0.05in}
The base ALM is a general purpose zero-shot model. The adaptation of ALM to a specific domain, should not lead to a drastic drop in its zero-shot performance. To verify if this holds for our method, we stup an ablation study. We adapt the base ALM CLAP on a target domain and check performance on other domains. For example, we adapt CLAP to ESC50 (dataset 1 in Table \ref{table: encoder results}) and check performance on remaining datasets (dataset \{2 - 12\} in Table \ref{table: encoder results}). This process can be repeated $T$ times, where $T$ is the number of target domains, to provide an estimate of drop in zero-shot performance. The numerical results of the experiments are shown in Table \ref{appendix: cross-domain generalization}. The diagonal entries (higlighted in blue), correspond to adaptation performed using one unlabelled audio sampled from the target domain data. 

\begin{figure}[!ht]
\centering
\includegraphics[width=0.461\textwidth, trim={0.25cm 0.12cm 0.22cm 0.25cm},clip]{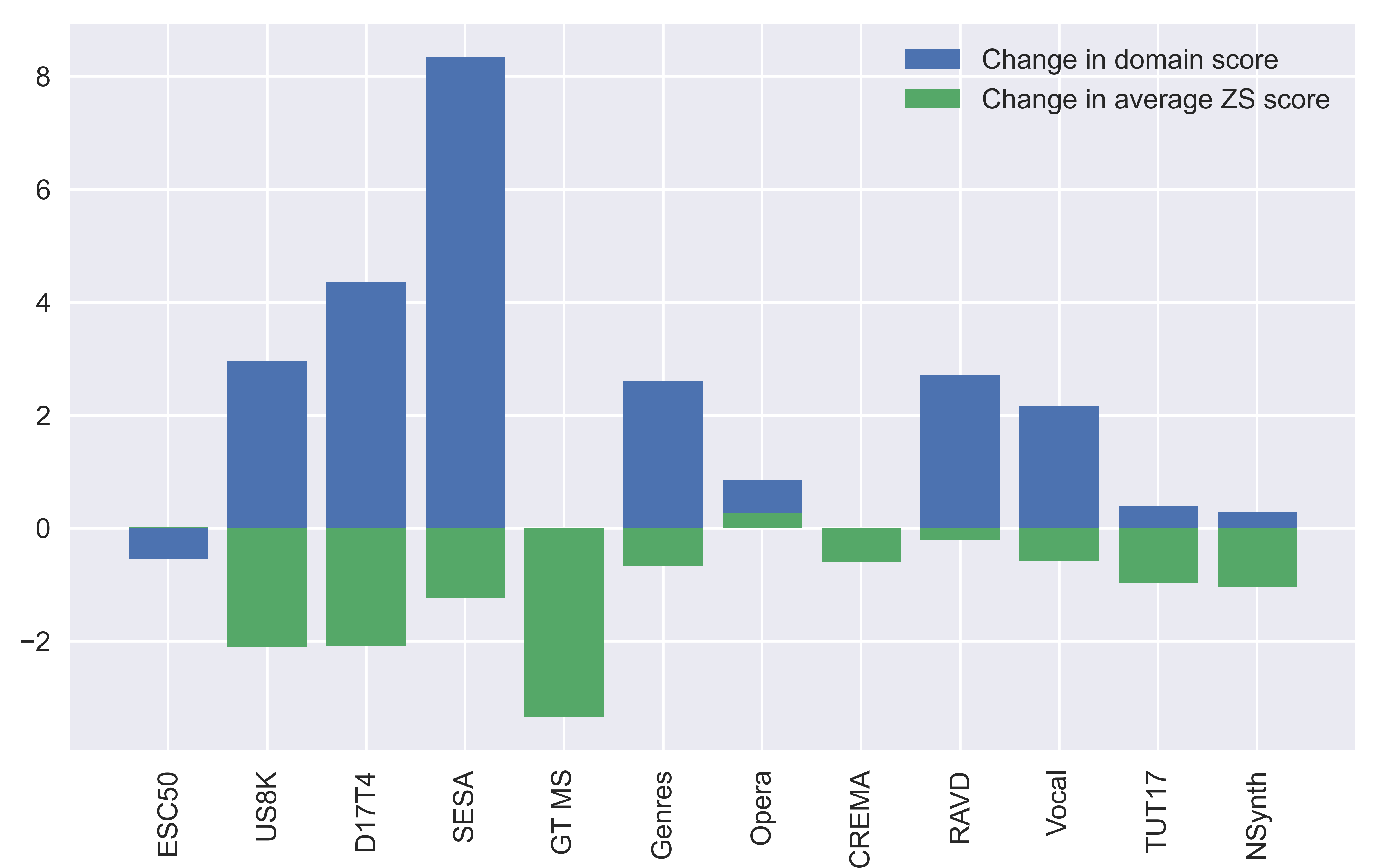}
\caption{Cross-domain adaptation performance. The x-axis indicates the dataset used for domain adaptation. The upper and lower bar indicate the change in - domain performance and average zero-shot score respectively.\vspace{-0.22in}}
\label{fig:cross}
\end{figure}

As we are interested in drop in generalization of ALM compared to the gain in the target domain performance, we summarize the findings from Table \ref{appendix: cross-domain generalization} in Fig \ref{fig:cross}. On average, we observe a 3\% increase in domain performance for an average zero-shot performance drop of 1\%. The exception is domain adpatation performed on ESC50, which slightly improves the  average zero-shot performance. 

\vspace{-0.1in}
\subsection{Limitations} \label{results: limitations}
Our adaptation method improves zero-shot performance by using one unlabelled audio at test-time. Its practical applications include on-device updates for audio-language model model to adapt it to the user’s audio and media. In this setup, access to labeled annotations is not possible but the model does have access to test audio used for inference. However, compared to zero-shot inference, the method incurs a higher computational cost because of backpropagation used to update the domain vector and the processing of multiple augmented views. Depending on the downstream application and the on-device compute, the parameter update may not be feasible, limiting the applicability of this method. 

\vspace{-0.2in}
\section{Conclusion} \label{sec: conclusion}
We propose a domain adaptation method for ALMs. The method uses unlabelled audio at test-time to enforce consistent prediction across augmented views of audio by optimizing a domain vector. It consists of three parts: Augment, Combine, and Optimise. Augment: The test audio is subject to various augmentations to simulate different listening conditions. Alongside this, a domain vector is incorporated into the model's text processing component; b) Combine: The model makes predictions based on the augmented audio and the modified text embeddings. These predictions are then averaged to produce a more stable and generalized output, and c) Optimize: Finally, the averaged prediction probabilities are used to compute self-entropy to fine-tune the domain vector, ensuring that the model's predictions are adapted for the domain. Adaption with one unlabelled audio provides an average 3.2\% (max 8.4\%) zero-shot improvement and with 5 audios an average improvement of 4.7\% (max 9.4\%). After adaptation, the ALM still retains its zero-shot generalization. 

\nocite{zhou2021narle,dhamyal2022describing,Qwen-Audio,Wang2023LauraGPTLA,alayrac2022flamingo,sgem,9413584}

\bibliographystyle{IEEEtran}
\bibliography{custom}

\begin{thebibliography}{10}
\providecommand{\url}[1]{#1}
\csname url@samestyle\endcsname
\providecommand{\newblock}{\relax}
\providecommand{\bibinfo}[2]{#2}
\providecommand{\BIBentrySTDinterwordspacing}{\spaceskip=0pt\relax}
\providecommand{\BIBentryALTinterwordstretchfactor}{4}
\providecommand{\BIBentryALTinterwordspacing}{\spaceskip=\fontdimen2\font plus
\BIBentryALTinterwordstretchfactor\fontdimen3\font minus \fontdimen4\font\relax}
\providecommand{\BIBforeignlanguage}[2]{{%
\expandafter\ifx\csname l@#1\endcsname\relax
\typeout{** WARNING: IEEEtran.bst: No hyphenation pattern has been}%
\typeout{** loaded for the language `#1'. Using the pattern for}%
\typeout{** the default language instead.}%
\else
\language=\csname l@#1\endcsname
\fi
#2}}
\providecommand{\BIBdecl}{\relax}
\BIBdecl

\bibitem{Elizalde2023NaturalLS}
B.~Elizalde, S.~Deshmukh, and H.~Wang, ``Natural language supervision for general-purpose audio representations,'' \emph{arXiv preprint arXiv:2309.05767}, 2023.

\bibitem{wu2022large}
Y.~Wu, K.~Chen, T.~Zhang, Y.~Hui, T.~Berg-Kirkpatrick, and S.~Dubnov, ``Large-scale contrastive language-audio pretraining with feature fusion and keyword-to-caption augmentation,'' \emph{arXiv preprint arXiv:2211.06687}, 2022.

\bibitem{pengi}
S.~Deshmukh, B.~Elizalde, R.~Singh, and H.~Wang, ``Pengi: An audio language model for audio tasks,'' \emph{arXiv preprint arXiv:2305.11834}, 2023.

\bibitem{gong2023listen}
Y.~Gong, H.~Luo, A.~H. Liu, L.~Karlinsky, and J.~Glass, ``Listen, think, and understand,'' \emph{arXiv preprint arXiv:2305.10790}, 2023.

\bibitem{mei2023wavcaps}
X.~Mei, C.~Meng, H.~Liu, Q.~Kong, T.~Ko, C.~Zhao, M.~D. Plumbley, Y.~Zou, and W.~Wang, ``Wavcaps: A chatgpt-assisted weakly-labelled audio captioning dataset for audio-language multimodal research,'' \emph{arXiv preprint arXiv:2303.17395}, 2023.

\bibitem{elizalde2022clap}
B.~Elizalde, S.~Deshmukh, M.~A. Ismail, and H.~Wang, ``Clap learning audio concepts from natural language supervision,'' in \emph{ICASSP 2023 - 2023 IEEE International Conference on Acoustics, Speech and Signal Processing (ICASSP)}, 2023, pp. 1--5.

\bibitem{li2023audio}
Y.~Li, X.~Wang, and H.~Liu, ``Audio-free prompt tuning for language-audio models,'' \emph{arXiv preprint arXiv:2309.08357}, 2023.

\bibitem{liang2023adapting}
J.~Liang, X.~Liu, H.~Liu, H.~Phan, E.~Benetos, M.~D. Plumbley, and W.~Wang, ``Adapting language-audio models as few-shot audio learners,'' \emph{arXiv preprint arXiv:2305.17719}, 2023.

\bibitem{10097117}
H.-H. Wu, O.~Nieto, J.~P. Bello, and J.~Salamon, ``Audio-text models do not yet leverage natural language,'' in \emph{ICASSP 2023 - 2023 IEEE International Conference on Acoustics, Speech and Signal Processing (ICASSP)}, 2023, pp. 1--5.

\bibitem{sun2020test}
Y.~Sun, X.~Wang, Z.~Liu, J.~Miller, A.~Efros, and M.~Hardt, ``Test-time training with self-supervision for generalization under distribution shifts,'' in \emph{International conference on machine learning}.\hskip 1em plus 0.5em minus 0.4em\relax PMLR, 2020, pp. 9229--9248.

\bibitem{zhou2022learning}
K.~Zhou, J.~Yang, C.~C. Loy, and Z.~Liu, ``Learning to prompt for vision-language models,'' \emph{International Journal of Computer Vision}, vol. 130, no.~9, pp. 2337--2348, 2022.

\bibitem{NEURIPS2022_bcdec1c2}
\BIBentryALTinterwordspacing
Y.~Gandelsman, Y.~Sun, X.~Chen, and A.~Efros, ``Test-time training with masked autoencoders,'' in \emph{Advances in Neural Information Processing Systems}, S.~Koyejo, S.~Mohamed, A.~Agarwal, D.~Belgrave, K.~Cho, and A.~Oh, Eds., vol.~35.\hskip 1em plus 0.5em minus 0.4em\relax Curran Associates, Inc., 2022, pp. 29\,374--29\,385. [Online]. Available: \url{https://proceedings.neurips.cc/paper_files/paper/2022/file/bcdec1c2d60f94a93b6e36f937aa0530-Paper-Conference.pdf}
\BIBentrySTDinterwordspacing

\bibitem{wang2021tent}
\BIBentryALTinterwordspacing
D.~Wang, E.~Shelhamer, S.~Liu, B.~Olshausen, and T.~Darrell, ``Tent: Fully test-time adaptation by entropy minimization,'' in \emph{International Conference on Learning Representations}, 2021. [Online]. Available: \url{https://openreview.net/forum?id=uXl3bZLkr3c}
\BIBentrySTDinterwordspacing

\bibitem{shu2022test}
M.~Shu, W.~Nie, D.-A. Huang, Z.~Yu, T.~Goldstein, A.~Anandkumar, and C.~Xiao, ``Test-time prompt tuning for zero-shot generalization in vision-language models,'' \emph{Advances in Neural Information Processing Systems}, vol.~35, pp. 14\,274--14\,289, 2022.

\bibitem{radford2021learning}
A.~Radford, J.~W. Kim, C.~Hallacy, A.~Ramesh \emph{et~al.}, ``Learning transferable visual models from natural language supervision,'' in \emph{International Conference on Machine Learning}.\hskip 1em plus 0.5em minus 0.4em\relax PMLR, 2021.

\bibitem{deshmukh2024pam}
S.~Deshmukh, D.~Alharthi, B.~Elizalde, H.~Gamper, M.~A. Ismail, R.~Singh, B.~Raj, and H.~Wang, ``Pam: Prompting audio-language models for audio quality assessment,'' \emph{arXiv preprint arXiv:2402.00282}, 2024.

\bibitem{park2019specaugment}
D.~S. Park, W.~Chan, Y.~Zhang, C.-C. Chiu, B.~Zoph, E.~D. Cubuk, and Q.~V. Le, ``Specaugment: A simple data augmentation method for automatic speech recognition,'' \emph{Interspeech 2019}, 2019.

\bibitem{wu2023audio}
H.-H. Wu, O.~Nieto, J.~P. Bello, and J.~Salomon, ``Audio-text models do not yet leverage natural language,'' in \emph{ICASSP 2023-2023 IEEE International Conference on Acoustics, Speech and Signal Processing (ICASSP)}.\hskip 1em plus 0.5em minus 0.4em\relax IEEE, 2023, pp. 1--5.

\bibitem{turian2022hear}
J.~Turian, J.~Shier \emph{et~al.}, ``{HEAR: Holistic Evaluation of Audio Representations},'' in \emph{NeurIPS 2021 Competitions and Demonstrations Track}, 2022.

\bibitem{esc50}
K.~J. Piczak, ``{ESC}: {Dataset} for {Environmental Sound Classification},'' in \emph{Proceedings of the 23rd {Annual ACM Conference} on {Multimedia}}.\hskip 1em plus 0.5em minus 0.4em\relax {ACM Press}, 2015, pp. 1015--1018.

\bibitem{UrbanSound}
J.~Salamon, C.~Jacoby \emph{et~al.}, ``A dataset and taxonomy for urban sound research,'' in \emph{22nd ACM international conference on Multimedia}, 2014.

\bibitem{mesaros2017dcase}
A.~Mesaros, T.~Heittola, A.~Diment, B.~Elizalde, A.~Shah, E.~Vincent, B.~Raj, and T.~Virtanen, ``Dcase 2017 challenge setup: Tasks, datasets and baseline system,'' in \emph{DCASE 2017-Workshop on Detection and Classification of Acoustic Scenes and Events}, 2017.

\bibitem{tzanetakis_essl_cook_2001}
\BIBentryALTinterwordspacing
G.~Tzanetakis, G.~Essl, and P.~Cook, ``Automatic musical genre classification of audio signals,'' 2001. [Online]. Available: \url{http://ismir2001.ismir.net/pdf/tzanetakis.pdf}
\BIBentrySTDinterwordspacing

\bibitem{cao2014crema}
H.~Cao, D.~G. Cooper, M.~K. Keutmann, R.~C. Gur, A.~Nenkova, and R.~Verma, ``{CREMA-D}: Crowd-sourced emotional multimodal actors dataset,'' \emph{IEEE transactions on affective computing}, vol.~5, no.~4, pp. 377--390, 2014.

\bibitem{ravdess}
\BIBentryALTinterwordspacing
S.~R. Livingstone and F.~A. Russo, ``The ryerson audio-visual database of emotional speech and song (ravdess): A dynamic, multimodal set of facial and vocal expressions in north american english,'' \emph{PLOS ONE}, vol.~13, no.~5, pp. 1--35, 05 2018. [Online]. Available: \url{https://doi.org/10.1371/journal.pone.0196391}
\BIBentrySTDinterwordspacing

\bibitem{vocalsound}
Y.~Gong, J.~Yu, and J.~Glass, ``Vocalsound: A dataset for improving human vocal sounds recognition,'' in \emph{IEEE International Conference on Acoustics, Speech and Signal Processing (ICASSP)}, 2022.

\bibitem{chen2022hts}
K.~Chen, X.~Du, B.~Zhu, Z.~Ma, T.~Berg-Kirkpatrick, and S.~Dubnov, ``Hts-at: A hierarchical token-semantic audio transformer for sound classification and detection,'' in \emph{ICASSP 2022-2022 IEEE International Conference on Acoustics, Speech and Signal Processing (ICASSP)}, 2022.

\bibitem{radford2019language}
A.~Radford, J.~Wu, R.~Child, D.~Luan, D.~Amodei, I.~Sutskever \emph{et~al.}, ``Language models are unsupervised multitask learners,'' \emph{OpenAI blog}, vol.~1, no.~8, p.~9, 2019.

\bibitem{ac3training}
S.~Deshmukh, B.~Elizalde, D.~Emmanouilidou, B.~Raj, R.~Singh, and H.~Wang, ``Training audio captioning models without audio,'' \emph{arXiv preprint arXiv:2309.07372}, 2023.

\bibitem{adam}
\BIBentryALTinterwordspacing
D.~P. Kingma and J.~Ba, ``Adam: A method for stochastic optimization,'' in \emph{ICLR (Poster)}, 2015. [Online]. Available: \url{http://arxiv.org/abs/1412.6980}
\BIBentrySTDinterwordspacing

\bibitem{zhou2021narle}
R.~Zhou, S.~Deshmukh, J.~Greer, and C.~Lee, ``Narle: Natural language models using reinforcement learning with emotion feedback,'' \emph{arXiv preprint arXiv:2110.02148}, 2021.

\bibitem{dhamyal2022describing}
H.~Dhamyal, B.~Elizalde, S.~Deshmukh, H.~Wang, B.~Raj, and R.~Singh, ``Describing emotions with acoustic property prompts for speech emotion recognition,'' \emph{arXiv preprint arXiv:2211.07737}, 2022.

\bibitem{Qwen-Audio}
Y.~Chu, J.~Xu, X.~Zhou, Q.~Yang, S.~Zhang, Z.~Yan, C.~Zhou, and J.~Zhou, ``Qwen-audio: Advancing universal audio understanding via unified large-scale audio-language models,'' \emph{arXiv preprint arXiv:2311.07919}, 2023.

\bibitem{Wang2023LauraGPTLA}
J.~Wang, Z.~Du, Q.~Chen, Y.~Chu, Z.~Gao, Z.~Li, K.~Hu, X.~Zhou, J.~Xu, Z.~Ma, W.~Wang, S.~Zheng, C.~Zhou, Z.~Yan, and S.~Zhang, ``Lauragpt: Listen, attend, understand, and regenerate audio with gpt,'' \emph{arXiv preprint arXiv:2310.04673}, 2023.

\bibitem{alayrac2022flamingo}
J.-B. Alayrac, J.~Donahue, P.~Luc, A.~Miech, I.~Barr, Y.~Hasson, K.~Lenc, A.~Mensch, K.~Millican, M.~Reynolds \emph{et~al.}, ``Flamingo: a visual language model for few-shot learning,'' \emph{Advances in Neural Information Processing Systems}, vol.~35, pp. 23\,716--23\,736, 2022.

\bibitem{sgem}
C.~Kim, J.~Park, H.~Shim, and E.~Yang, ``{SGEM}: Test-time adaptation for automatic speech recognition via sequential-level generalized entropy minimization,'' in \emph{Conference of the International Speech Communication Association (INTERSPEECH)}, 2023.

\bibitem{9413584}
Y.~Wang, N.~J. Bryan, M.~Cartwright, J.~Pablo~Bello, and J.~Salamon, ``Few-shot continual learning for audio classification,'' in \emph{ICASSP 2021 - 2021 IEEE International Conference on Acoustics, Speech and Signal Processing (ICASSP)}, 2021, pp. 321--325.

\end{thebibliography}


\end{document}